\documentclass[12pt,preprint]{aastex}

\shortauthors{Xu, Liu \& Liu}

\usepackage{emulateapj5} 
\usepackage{color}

\begin{document}

\title{A likely Micro-quasar in the Shadow of M82 X-1}

\author{Xiao-jie Xu\altaffilmark{1,2}, Jifeng Liu\altaffilmark{2} and Jiren Liu\altaffilmark{2}}
\altaffiltext{1}{Department of Astronomy and Key Laboratory of Modern Astronomy and Astrophysics, Nanjing University, Nanjing, P. R. China 210093}
\altaffiltext{2}{Key Laboratory of Optical Astronomy, National Astronomical Observatories of China, 20 Datun Rd, Chaoyang, Beijing, China 100012}

\email{xuxj@nju.edu.cn}
\email{jfliu@nao.cas.cn}

\begin{abstract}

The	ultra-luminous X-ray source (ULX) M82 X-1 is one of the most promising intermediate mass black hole candidates in
the local universe based on its high X-ray luminosities ($10^{40}-10^{41}
{\rm~erg~s^{-1}}$) and quasi-periodic oscillations, and is possibly associated
with a radio flare source.
In this work, applying the sub-pixel technique to the 120 ks Chandra
observation (ID: 10543) of M82 X-1, we split M82 X-1 into two sources separated by 1.1$\arcsec$. The secondary source is not detected in other M82 observations. 
The radio flare source is found to associate not with M82 X-1, but with the nearby
transient source S1 with an outburst luminosity of $\sim 10^{39}
{\rm~erg~s^{-1}}$.
With X-ray outburst and radio flare activities analogous to the recently
discovered micro-quasar in M31, S1 is likely to be a micro-quasar hidden in the
shadow of M82 X-1.

\end{abstract}

\keywords{Galaxy: individual(M 82) --- X-rays: binaries}

\section{INTRODUCTION}
The ULX M82 X-1, with X-ray luminosities of $\gtrsim 10^{40}{\rm~erg~s^{-1}}$ and quasi-periodical oscillation (QPO) behavior, is one of the
most promising intermediate mass black hole (IMBH) candidates in the local
universe \citep{col94,gri00,mat01,str03,pas14}. 
The discovery of X-ray spectral state transition in M82 X-1 shows it has X-ray
luminosities  of $L_{\rm X}~\sim~10^{41}~{\rm~erg~s^{-1}}$ at the
thermal-dominant state, suggesting an IMBH with mass $\sim 200$ to $800
M_{\odot}$ \citep{feng10}. 
The discovery of the 60-120 milli-Hz QPOs
\citep{str03,fio04,muc06,cab13,pas13a} suggests the strong X-ray emission
originates from the accretion disk instead of a jet, and suggests an IMBH of
$\sim2\times10^4M_\odot$ if these low-frequency QPOs corresponds to the
Keplerian frequency of the innermost stable orbit.  
Pasham et al. (2014) re-analyzed 6-year RXTE X-ray observations and revealed
the high-frequency, 3:2 ratio-ed, twin-peak QPOs of 3.3Hz and 5Hz, which, in
combination with the low-frequency QPO revealed by an XMM-Newton observation,
suggests an IMBH of $415\pm63M_\odot$ under the relativistic precession model
(Motta et al. 2013). 

Many efforts have been made to search for the radio counterpart of M82 X-1, and
two radio sources were discovered at the vicinity of this ULX. 
A stable radio source $0.8\arcsec$ away from M82 X-1 is proposed to be a
supernova remnant (SNR) and not related to M82 X-1
\citep{col94,mcd02,kor05,gen13}. 
The other radio source was detected (at R.A.=9:55:50.19, Dec=+69:40:46.0)
throughout the 7-hr 1981 VLA observation at a 9mJy flux level \citep{kro85},
but later was not detected in the 2003 VLA observations \citet{kor05} or the
2005 VLA observations \citep{kaa06b}.
\citet{kor05} re-examined the 1981 VLA data and found a $0.5\arcsec$ offset
between the flare source and M82 X-1.  Given the Chandra positional uncertainty
of $0.6\arcsec$, it is not clear whether it is really associated with M82 X-1
and suggesting relativistic beaming effects as causing the high X-ray
luminosities. A careful work on astrometry is required to achieve a better
understanding of M82 X-1. 

In this paper, we re-analyze the \textit{Chandra} ACIS observations with the sub-pixel
technique to obtain improved spatial resolution and
centroid accuracy. We describe data analysis and results
in \S~2 and discuss the implications of our work in \S~3. 

\section{DATA ANALYSIS \& RESULTS}

\subsection{Spatial Analysis}
\begin{figure}
\centering
\includegraphics[height=3in,width=4.4in]{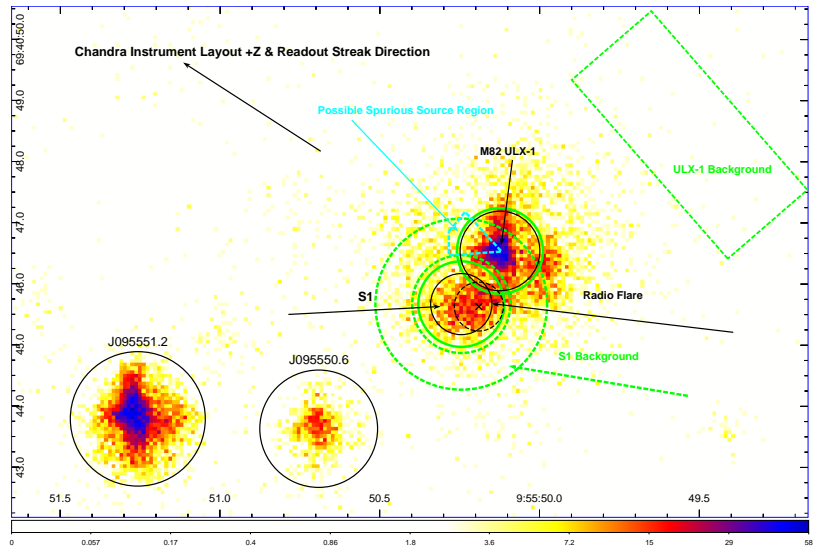}
   \caption{1/8 pixel \textit{Chandra} 0.3-8 keV image of M82 X-1. Solid black circles are 3-$\sigma$ error circles for detected sources as noted. Solid green circles, dashed green annulus and rectangle are for source and background spectra extraction. The \textit{Chandra} instruments $+Z$ and readout streak directions are given at the upper-left corner. The cyan region shows the possible spurious source region caused by the PSF distortion. The `X' and the dashed black circle mark the radio flare position and its uncertainty brought by astrometry correction, respectively. The two sources J095551.2 and J095550.6 marked with solid black circles at the lower-left corner are two nearby bright X-ray sources. Note that there are no count excess in the lower left direction of J095551.2.}
   \label{Fig:01}
\end{figure}

M82 X-1 has been observed 22 times by \textit{ACIS} onboard \textit{Chandra} between 1999 and 2014, with
off-axis angles from below $1^\prime$ to above $4^\prime$. 
The Obs IDs are: 361, 378, 379, 380, 1302, 2933, 5644, 6097, 6361, 8190, 10025, 10026, 10027, 10542, 10543, 10544, 10545, 10925, 11104, 11800, 13796 and 15616. In this work we focus on Obs ID 10543 (PI: Strickland), which had the longest
exposure of 120 ks with  M82 X-1 placed only $\sim~1^\prime$ from the aiming
point. The energy range used in this paper is 0.3-8 keV unless stated otherwise. A careful examination of the Chandra image shows that the image size of
M82 X-1 is larger than other nearby sources (e.g., J095551.2, see Fig~\ref{Fig:01}), motivating our further analysis. 

We reprocess the Level 1 event files using the \textit{chandra\_repro} command
in \textit{Chandra} Interactive Analysis of Observations (\textit{CIAO},version
4.5) software. Following the standard pipeline\footnote{see
http://cxc.harvard.edu/ciao/ for details}, we remove time intervals with
significant flares, as well as bad columns/pixels and events of bad grades, to
produce the Level 2 event files.  The \textit{check\_vf\_pha} parameter was set
to {\textit no} to avoid false-removing of good events. The \textit{sub\_pixel}
parameter was set to \textit {EDSER} to enable sub-pixel analysis. 
Fig~\ref{Fig:01} shows the 1/8 pixel \textit{ACIS} 0.3-8 keV image centered
around M82 X-1 after reprocessing. Clearly, M82 X-1 is split into two fuzzy
but discrete sources, while other nearby ULXs are not. Applying wavdetect to
the image results in an detection of two sources, as presented in
Table~\ref{tbl-01} and Fig~\ref{Fig:01}. The brighter one is M82 X-1 itself,
and its position (RA=9:55:50.123, Dec=+69:40:46.54) is consistent with
previously detected positions (e.g., Griffiths et al. 2000). 
The second source (hereafter S1), with RA=9:55:50.245 and Dec=+69:40:45.67, is
located $1.1\arcsec$ to the lower-left of M82 X-1.
Not detected by wavdetect is extra photon concentration to the lower right of X-1, which is probably caused by the PSF feature like  other nearby ULXs (e.g., see J095551.2 in Fig~\ref{Fig:01}).
In ObsID 10543, M82 X-1 suffers from severe pile-up effects (see \S2.2), which may affect
its centroid.
In comparison, the \textit{HRC}
observations do not suffer from pile-up effects and are expected to give accurate 
centroid positions for M82 X-1 and other sources.
We have analyzed the \textit{HRC} observations and listed the centroid positions
for matched sources between the ACIS and HRC (Obs ID 8189) observations.
As shown in Table~\ref{tbl-02}, similar displacement of $\sim0.6$\arcsec\ exists
between the ACIS and HRC observations for all matched pairs, including X-1 despite 
the pile-up effects.

As \textit{Chandra}'s PSF could be strongly distorted and cause fake
detection in $+Z$ direction (\textit{Chandra} instrument layout directions are used in this paper for convenience)\footnote{see
http://cxc.harvard.edu/cal/Hrc/PSF/acis\_psf\_2010oct.html and
http://cxc.cfa.harvard.edu/ciao/caveats/psf\_artifact.html for details}, we
need to ascertain whether they are real sources or just PSF artifacts.

The relative position of S1 strongly suggested S1 is a real source.
The relative position of S1 to M82 X-1 in CCD is $-Y$, which is different from the possible spurious source region (Fig~\ref{Fig:01}, the cyan region
centered at 15$^{\circ}$ counter clockwise to $+Z$) or the readout streak region (Fig~\ref{Fig:01}). Moreover, the distance from X-1 to
S1 is $\sim$1.1\arcsec, greater than the PSF distortion
region($\sim$0.8\arcsec), making it unlikely to be a PSF artifact.
Indeed, none of other nearby bright sources show photon concentration in $-Y$
direction like S1, again supporting that S1 is a real source. 

\subsection{Spectral Analysis}
The spectra and the respective \textit{rmf} and \textit{arf} files of M82 X-1
and S1 are extracted using \textrm{specextract} command in \textit{CIAO} with
0.7\arcsec\ circular source regions (Fig~\ref{Fig:01}).  The background region
for X-1 is a rectangular region, and that for S1 is an annulus centered at S1,
with inner radius 0.8\arcsec\ and outer radius 1.4\arcsec.  The X-1 region is
excluded when extracting spectrum for S1, and vice versa. 
The respective count rates are roughly 0.029 c/s and 0.016 c/s in 0.3-8 keV band for X-1 and S1.
The 0.7\arcsec\ extraction radius of S1 may miss certain amount of hard
photons outside, thus lead to a biased spectrum. Yet our results are
qualitatively valid due to the following reasons: i) the 0.7\arcsec\ radius is
comparable to the local $\sim$85\% energy encircled radius as suggested by the
PSF file, so the number of leaked hard photons is limited; ii) further
correction on the spectrum will include more hard photons and thus a harder
spectrum.
The spectra are then loaded to {\rm xspec} and analyzed. We do not tend to
focus on the spectra investigation, so an absorbed power law (pha*powerlaw)
model is used as a characterization. 

As shown in Table~\ref{tbl-01}, the spectrum of M82 X-1 has $\Gamma_{\rm ULX}
\sim -0.7$, even harder than the one reported by \citet{agr06}. Such a hard
spectrum is unphysical and likely caused by heavy pile-up effects, as indicated
by the strong read-out streak.
Because the readout streak contains photons of M82 X-1 before pile-up, its
spectrum (with a 0.3-8 keV count rate of 0.009 c/s) is analyzed as
a proxy for M82 X-1.
The spectrum of the readout streak can be well fitted with an absorbed power law with
$\Gamma_{\rm streak} = 1.85^{+0.35}_{-0.33}$ with $\chi_{\nu}^2/d.o.f. =
0.64/62$. 
The streak exposure time correction is found to be $\sim240$, suggesting that
the true count rate of M82 X-1 is 2.1 c/s, and its un-absorbed luminosity is
$8.7\times 10^{40} {\rm~erg~s^{-1}}$.

The spectrum of S1 has $NH_{\rm S1} =
1.08^{+0.27}_{-0.24}\times10^{22}$cm$^{-2}$ and $\Gamma_{\rm S1} =
1.12^{+0.22}_{-0.21}$ with $\chi_{\nu}^2/d.o.f. = 1.38/48$. Placed at the distance of M82, S1 would have an un-absorbed luminosity of
$7.2\times 10^{38} {\rm~erg~s^{-1}}$, $\sim$ 1\% of that of M82 X-1.
However, due to the severe pile-up of M82 X-1, the 0.3-8 keV count rate of S1 is almost
half of M82 X-1's count rate, enough to cause a significant positional
displacement of M82 X-1.

To test the spectral difference for subregions of the PSF, we have extracted
the spectra from two subregions of the nearby bright source J095551.2 in
parallel to X-1 and S1 of M82 X-1 (see Fig~\ref{Fig:01}). 
As expected, these spectra are similar to each other with $\Gamma \sim 2.0$.

As shown above, S1's spectrum is consistent with those in the low-hard states of black hole
X-ray binaries (Remillard \& McClintock 2006) and is intrinsically different
from that of M82 X-1 or its readout streak at the 90\% confidence level, again supporting that S1 is a real source. 
\subsection{Temporal Behavior of S1}
\begin{figure}
\centering
\includegraphics[height=3in,width=4.4in]{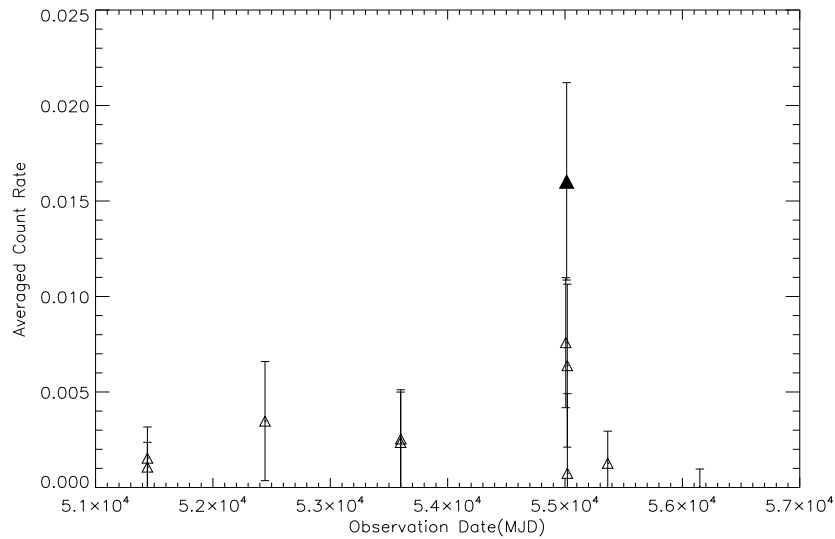}
   \caption{Averaged 0.3-8 keV net count rates of S1 in \textit{Chandra ACIS} observations between September 1999 to February 2013. The 12 points from left to right refer to Obs ID 361, 1302, 2933, 5644, 6361, 10542, 10543, 10925, 10544, 11104, 13796 and 15616, respectively. The filled triangle corresponds to the detection of S1 in Obs ID 10543, and the opened triangles are from observations without detection and should be treated as upper limits.}
   \label{Fig:02}
\end{figure}

We further analyze other \textit{Chandra ACIS} observations in a similar way as in Obs ID 10543. S1 is not detected in any other \textit{ACIS} observations, including the three observations 10542, 10925 and 10544, observed about a week before and after Obs ID 10543, respectively. Using the same 0.7$\arcsec$ extraction region and the same annular background region, we extract counts at S1's position for all other \textit{ACIS} observations. In Obs IDs 378, 379, 380, 6097, 8190, 10025, 10026, 10027 and 10545, S1 was fully buried in the PSF wings of X-1, so extraction results from these observations are excluded. In other observations, the PSF wing of M82 X-1 still partly cover S1 regions, and these obvious X-1 PSF wing regions are excluded. After this process, it is still possible that some photons in the extraction regions could come from X-1, thus the estimated count rates of S1 should be treated as upper limits. In Fig~\ref{Fig:02} we present the results. The 0.3-8 keV count rate of S1 was $\lesssim$ 0.001 c/s in 1999 (Obs ID 361) and was $\lesssim$ 0.007 c/s in 2009 (Obs ID 10542), then rose to $\sim$ 0.016 c/s within a week (Obs ID 10543) and dropped to $\lesssim$0.0007 c/s with 7 days (Obs ID 10544). Judged from the light curve, S1 is likely a transient source in 0.3-8 keV band. This transient behavior, as well as the fact that S1 was usually in the shadow of M82 X-1's PSF is likely the reason why S1 was only detected once in 22 Chandra observations.
\subsection{Astrometry}
It is now possible to identify the radio flare source given the centroids of
M82 X-1 and S1.
Using three bright ACIS point sources ($\sigma>60$; see Table~\ref{tbl-02} for
details) with optical matches from the Sloan Digital Sky
Survey\citep[SDSS,][]{ahn14}, we find a $\sim0.4\arcsec$ displacement between
the ACIS coordinate frame and the SDSS coordinate frame (the same as the VLBI
coordinate frame with an error of $0.2\arcsec$). 
The position of the radio flare source is corrected
accordingly and registered to the ACIS image (Fig~\ref{Fig:01}).
The flare source is now $0.96\arcsec$ away from M82 X-1, but only $0.28\arcsec$
away from S1.
Clearly, the position of M82 X-1 is no longer consistent with that of the radio
flare source. 
On the other hand, S1's position is consistent with the radio flare source
within the positional uncertainty ($\sim0.4\arcsec$; mainly from the
astrometric correction). 

If the radio flare is indeed associated with S1, it can be the radio outburst
from a black hole X-ray binary when the binary transits from hard to soft states
\citep[e.g.,][]{fen04}. 
Recently, \citet{mid13} discovered strong radio emission at a flux level
$\leq1$mJy from a micro-quasar in M31 during its X-ray outburst at an X-ray
luminosity of $1.3\times10^{39} {\rm~erg~s^{-1}}$. 
The properties of S1 are consistent with those of the M31 micro-quasar, and it
can well be a similar micro-quasar, albeit with higher radio flux (9mJy).
However, unlike in the case of the M31 micro-quasar, the radio flare and the
X-ray outburst for S1 were not detected simultaneously, and further
simultaneous monitoring observations in radio and X-ray are needed to determine
its exact nature.

\section{DISCUSSION}

M82 X-1, one of the most promising IMBH candidates in the local universe, is
possibly associated with a radio flare source, suggestive of relativistic
beaming effects for its extremely high X-ray luminosities.
We have applied the sub-pixeling technique to the longest Chandra
observation of M82 X-1, and split it into two separate discrete sources,
namely X-1 itself and S1.
Both relative positions and spectral properties show that S1 is a real source
and not a PSF artifact, although it was detected only once out of 22 Chandra
observations.
Careful astrometric work shows that the radio flare source is associated with
the transient S1. 
M82 X-1 is no longer associated with the radio flare, eliminating the necessity
for relativistic beaming.
The radio flare and X-ray outburst of S1 are analogous to those of a recently
discovered micro-quasar in M31 (Middleton et al. 2013), making it a likely
micro-quasar in the shadow of M82 X-1.

The discovery of S1, however, does not affect previous results on M82 X-1's
spectral properties significantly (e.g., Feng et al. 2010).  
This is because S1 luminosity, even in its outburst, is only $\sim 7\times10^{38}
{\rm~erg~s^{-1}}$. In comparison, M82 X-1 exhibited X-ray luminosities from
$2\times10^{40}$ to $\geq10^{41} {\rm~erg~s^{-1}}$ (e.g., Kaaret et al. 2006; Feng
et al. 2010), 30-100 times higher than the outburst luminosity of S1. 
Furthermore, S1 was detected only once out of 22 Chandra observations between
1999 and 2014, and it must remain in a quiescent state at even lower
luminosities most ($>$95\%) of the time .
However, S1 could become important for on-axis observations when M82 X-1
suffers severe pile-up.  In the case of Obs ID 10543, the count rate of S1 is
almost half of the observed count rate of M82 X-1.

\acknowledgements

The authors thank the anonymous referee for comments that helped improve the paper. The authors thank Drs. Zhiyuan Li and Junfeng Wang for help discussions on data analyzing methods. The authors acknowledge support from National Science Foundation of China through grants NSFC-11273028, NSFC-11333004 and NSFC-11303015.

\begin{table}[h]
\caption[]{Results of spatial and spectral analysis of M82 ULX-1, the readout streak and S1 in 0.3-8 keV. An absorbed power law model is used to characterize the spectra.}
\label{tbl-01}
\small
  \begin{center}\begin{tabular}{ccccccccc}
  \hline\hline
Source & RA& Dec&Net Counts&Significance&NH & $\Gamma$  & $\chi_{\nu}^2$/d.o.f. \\
&    &&     && ($1.0\times10^{22}$cm$^{-2}$)      &               &\\
  \hline\noalign{\smallskip}
ULX-1  & 9:55:50.123&+69:40:46.54&4165&238$\sigma$&0.56&-0.68 &2.1/174\\
ULX-1 Streak &-&- &-&-& $1.25^{+0.38}_{-0.33}$&$1.85^{+0.35}_{-0.33}$ &0.64/62\\
S1  & 9:55:50.245&+69:40:45.67&2190&71$\sigma$&$1.08^{+0.27}_{-0.24}$&$1.12^{+0.22}_{-0.21}$ &1.38/48\\

\hline
\end{tabular}
\end{center}
\end{table}

\begin{table}[h]
\caption[]{Positions of sources for ACIS-HRC and ACIS-SDSS astrometric registration.}
\label{tbl-02}
\tiny
  \begin{center}\begin{tabular}{ccccccc}
  \hline\hline
\multicolumn{7}{c}{Selected Sources Detected in Both ACIS and HRC Observations}\\
\hline
ACIS RA&ACIS Dec& Net Counts&Significance&HRC RA&HRC Dec & Displacement\\
\hline
9:55:50.123$^a$&+69:40:46.54&4165&238$\sigma$& 9:55:50.234&+69:40:46.27& 0.63\arcsec \\
9:55:50.689&+69:40:43.61&1744&103$\sigma$&9:55:50.803&+69:40:43.37& 0.63\arcsec \\
9:55:51.257&+69:40:43.79&7063&501$\sigma$& 9:55:51.395&+69:40:43.58 &0.73\arcsec \\
9:55:50.355&+69:40:36.36& 1747&86$\sigma$&9:55:50.473&+69:40:36.18& 0.63\arcsec \\
9:55:51.479&+69:40:35.90& 2252&105$\sigma$&9:55:51.591&+69:40:35.70& 0.60\arcsec \\
9:55:47.488&+69:40:59.61& 587&61$\sigma$&9:55:47.585&+69:40:59.42& 0.53\arcsec \\
9:55:14.583$^b$&+69:47:35.55&2201&131$\sigma$&9:55:14.496 &+69:47:35.34&0.49\arcsec\\
 \hline\hline
\multicolumn{7}{c}{Bright ACIS X-ray Sources identified in SDSS Observations}\\
\hline
ACIS RA&ACIS Dec& Net Counts&Significance&SDSS RA&SDSS Dec & Displacement\\
\hline
9:56:58.658&+69:38:52.13&  652&90$\sigma$  &9:56:58.678 &+69:38:52.54&0.422\arcsec\\
9:55:05.206&+69:44:42.47&  574 &61$\sigma$   &9:55:05.227 &+69:44:42.85&0.394\arcsec\\
9:55:14.583$^b$&+69:47:35.55&  2201&131$\sigma$   &9:55:14.546 &+69:47:35.85&0.354\arcsec\\

\hline
\end{tabular}
\end{center}
\tablecomments{\scriptsize 
$^a$ M82 ULX-1.\\
$^b$ The source detected in HRC, ACIS and SDSS observations.
}
\end{table}

\end{document}